\documentclass[sigconf]{acmart}

\AtBeginDocument{%
  }

\copyrightyear{2023}
\acmYear{2023}
\setcopyright{acmlicensed}\acmConference[KDD '23]{Proceedings of the 29th ACM SIGKDD Conference on Knowledge Discovery and Data Mining}{August 6--10, 2023}{Long Beach, CA, USA}
\acmBooktitle{Proceedings of the 29th ACM SIGKDD Conference on Knowledge Discovery and Data Mining (KDD '23), August 6--10, 2023, Long Beach, CA, USA}
\acmPrice{15.00}
\acmDOI{10.1145/3580305.3599796} 
\acmISBN{979-8-4007-0103-0/23/08}

\usepackage{hyperref}
\usepackage{latexsym}
\usepackage{mathrsfs}
\usepackage{multirow}
\usepackage{amsmath}
\usepackage{amsfonts}
\usepackage{subfigure}
\usepackage{graphicx}
\usepackage{algorithmic}
\usepackage{comment}
\usepackage{newtxmath}
\usepackage{balance}
\usepackage{multirow}
\usepackage{algorithm}

\begin{document}

\title{Controllable Multi-Objective Re-ranking with Policy Hypernetworks}

\author{Sirui Chen}
\authornote{These authors contributed equally to this work.}
\affiliation{
    \institution{School of Information, \\ Renmin University of China}
    \city{}
    \country{}
    \ chensr16@gmail.com
}

\author{Yuan Wang}
\authornotemark[1]
\authornote{First corresponding author: Xiao Zhang. Second corresponding author: Yuan Wang
}
\affiliation{
    \institution{Alibaba Group}
    \city{}
    \country{}
    \ wy175696@taobao.com
}
\author{Zijing Wen}
\affiliation{
    \institution{Alibaba Group}
    \city{}
    \country{}
    \ wzj267727@taobao.com
}
\author{Zhiyu Li}
\affiliation{
    \institution{Alibaba Group}
    \city{}
    \country{}
    \ tuanyu.lzy@taobao.com
}
\author{Changshuo Zhang}
\affiliation{
    \institution{Gaoling School of AI (GSAI) \\ Renmin University of China}
    \city{}
    \country{}
    \ lyingcs@foxmail.com
}
\author{Xiao Zhang}
\authornotemark[2]
\affiliation{
    \institution{Gaoling School of AI (GSAI) \\ Renmin University of China}
    \city{}
    \country{}
    \ zhangx89@ruc.edu.cn
}

\author{Quan Lin}
\affiliation{
    \institution{Alibaba Group}
    \city{}
    \country{}
    \ tieyi.lq@taobao.com
}
\author{Cheng Zhu}
\affiliation{
    \institution{Alibaba Group}
    \city{}
    \country{}
    \ xize.zc@taobao.com
}
\author{Jun Xu}
\affiliation{
    \institution{Gaoling School of AI (GSAI) \\ Renmin University of China}
    \city{}
    \country{}
    \ junxu@ruc.edu.cn
}

\renewcommand{\shortauthors}{Sirui Chen et al.}

\begin{abstract}
Multi-stage ranking pipelines have become widely used strategies in modern recommender systems, where the final stage aims to return a ranked list of items that balances a number of requirements such as accuracy, diversity etc. 
Typically, linear scalarization is used to merge these requirements into a single optimization objective by summing them with preference weights. 
However, existing final-stage ranking methods often rely on static models where preference weights are determined during offline training and remain unchanged during online serving. Adjusting these weights requires retraining the models, which can be time-consuming and resource-intensive. Moreover, the most suitable weights may vary significantly for different user groups or at different time periods, such as during holiday promotions.
In this paper, we propose a framework called controllable multi-objective re-ranking (CMR) which incorporates a hypernetwork to generate parameters for a re-ranking model according to different preference weights. In this way, CMR is enabled to adapt the preference weights according to the environment changes in an online manner, without any retraining. Moreover, we classify practical business-oriented tasks into four main categories and seamlessly incorporate them in a new proposed re-ranking model based on an Actor-Evaluator framework, which serves as a reliable real-world testbed for CMR. Offline experiments based on the dataset collected from Taobao App showed that CMR improved several popular re-ranking models by using them as underlying models. Online A/B tests also demonstrated the effectiveness and trustworthiness of CMR.
\end{abstract}



\maketitle
\section{Introduction}
Multi-stage strategies are prevalent in today's industrial recommender systems, featuring high scalability and low response latency, retrieving relevant items from billion candidates within tens of milliseconds. Early recommender systems~\cite{borisyuk2016casmos,yi2019sampling,zhao2019recommending} 
estimate the relevance of each item to a user and rank items according to their relevance scores. 
In recent years, recommender systems have placed more emphasis on other important objectives of the results, such as diversity~\cite{xia2017adapting}, fairness~\cite{oosterhuis2021computationally}, serendipity~\cite{kotkov2016survey}, and unbiasedness~\cite{Zhang2021Counterfactual,Zhang2022Counteracting} etc. It requires the system to be able to capture the relationship between items and balance multiple objectives, where re-ranking models excel. Therefore, the final-stage re-ranking model \cite{bello2018seq2slate, wilhelm2018practical} is gaining its popularity as a solid supplement of the traditional systems.

Linear scalarization~\cite{birge2011introduction} is the most widely used technique to merge multiple optimization objectives into a single one, by weighted summing the objectives. Existing multi-objective re-ranking models conduct the linear scalarization in a static way: the trade-off between tasks, or the set of preference weights, is predefined and fixed during the training and testing phases, resulting in a solution corresponding to a single preference combination. To obtain the performances of different preference weights, in principle, one needs to train the model multiple times, each for a single combination.

In real-world applications, however, dynamic adjustment of the preference weights without model retraining is strongly favored. First, it enables fast hyperparameter tuning. Even with prior knowledge, it is still not trivial to pick the most appreciated preference weights. Tuning methods like grid-search and trail-and-error can be greatly accelerated if repeated model training can be avoided. Second, it enables fast system response to environmental changes. Modern recommender systems such as Amazon need real-time flow control during Black Friday, adjusting their recommendation strategies based on users' behavior. Ideally, the adjustments should take effect immediately. In industrial applications, re-training a model takes hours or even days. Last but not the least, different groups of users may have diverse preferences. As a result, the best set of preference weights varies from group to group. Training multiple models for different user groups is often prohibitively expensive in industrial applications. Dynamic weight adaptation helps to serve diverse groups with a single model. 

To achieve online dynamic preference adjustment with a single model, we propose a new controllable multi-objective re-ranking framework (CMR). Inspired by recent studies on multi-task leaning~\cite{navon2020learning,raychaudhuri2022controllable,lin2020controllable}, CMR is designed to consist of a hypernetwork and a normal re-ranking model, where the hypernetwork generates special parameters of the re-ranking model for a given set of preference weights. With the generated parameters, the re-ranking model can produce suitable recommended lists for different preference weights. During the offline training, the preference weights are randomly sampled from a feasible distribution to simulate the different environment changes. The parameter generation policy of the hyper network and the re-ranking model are jointly optimized. During the online serving, the preference weights can be specified in real time as desired, without the need for model retraining. 

Besides, many business-oriented issues need to be considered in industrial e-commerce recommender applications. For example, to increase sellers’ engagement, a recommender system may place more seller-generated contents in results rather than auto-generated contents, even with a certain expense of lowering the relevance. To ensure the functional positioning of a scene, an e-commence recommender system may prioritize new contents instead of old ones. We conducted a practical review of real-life business-oriented tasks and classify them into four main types: fixed-position insertions, flow control, diversity, and group ordering. Many real-life complex business problems can be broken down into these tasks or their combinations. In traditional recommender systems, these tasks are handled separately in a pipeline fashion, leading to sub-optimal performance. Apart from simplicity and feasibility, it is mainly due to the absence of any joint optimal solution. In this paper, we devise a new re-ranking model based on the Actor-Evaluator (AE) framework which can decently incorporate all four types of business-oriented tasks, offering an end-to-end optimal solution. It also serves as a sound realistic test bed for CMR. 

We evaluated CMR with various re-ranking models and demonstrated its applicability across different models. The evaluations were conducted on a large-scale dataset from the Taobao app. Online experiments showcased the flexibility of CMR and the effectiveness of the proposed AE re-ranking model specifically for Taobao.

The main contributions of the paper are:\\
\textbf{(1) Problem and framework.} We propose the controllable multi-objective re-ranking framework (CMR) to address the online real-time preference adjustment of multi-objective modeling. It consists of a hyper network and a customizable re-ranking model.\\
\textbf{(2) A controllable re-ranking model for business-oriented tasks.} We classify real business-oriented tasks into four types and propose an AE-based re-ranking model that integrates all these task types. It provides an end-to-end optimal solution, replacing the previous pipeline approach and serving as a realistic test bed for CMR.\\
\textbf{(3) Evaluation.} Through extensive online and offline experiments, we validate the effectiveness and flexibility of CMR and demonstrate the capabilities of the proposed model on the Taobao App.

\section{Related Work}

\textit{Re-ranking} is the final stage of a Multi-stage Recommender System (MRS) that re-ranks top candidates to improve recommendation performance. Re-ranking models explicitly consider list-wise context, distinguishing them from earlier ranking stages. Early approaches used maximum marginal relevance to add items to the list~\cite{carbonell1998use}, while recent works have adopted deep Neural Networks (NN)~\cite{lecun2015deep,goldberg2017neural}. Re-ranking works can be classified based on objectives, such as accuracy~\cite{karypis2005item,koren2009matrix}, diversity~\cite{xia2016modeling,xia2017adapting,jiang2017learning,liu2020dvgan}, and fairness~\cite{karako2018using,oosterhuis2021computationally,zhu2021fairness,xu2023p}. Re-ranking models have two training paradigms: learning by observed signals and learning by counterfactual signals. Models trained by observed signals require specifying the best list as labels, which can be challenging due to the large list space~\cite{zhuang2018globally,bello2018seq2slate}. In contrast, models trained by counterfactual signals are based on the AE framework and evaluate lists instead of specifying the best list~\cite{huzhang2021aliexpress,gong2019exact,wang2019sequential}. Multi-task modeling based on the AE framework has not been well explored~\cite{liu2022neural}.

\textit{Multitask Learning (MTL)} seeks to learn a single model to simultaneously solve several learning problems while sharing information among tasks \cite{zhang2021survey,ruder2017overview}. In recent years, a variety of deep MTL networks with hard or soft parameter sharing structures have been proposed \cite{misra2016cross,long2017learning,yang2016deep}.
Another set of works treats MTL as a multi-objective optimization problem that aims to identify Pareto stationary solutions across various tasks \cite{lin2019pareto, mahapatra2020multi, xie2021personalized} and applied in recommender system \cite{jannach2022multi, li2020video, zheng2022survey}. These studies employ different approaches: some alternate between optimizing the joint loss and adjusting the weight of each loss \cite{lin2019pareto}, while others frame the dynamic weight optimization process as a personalized sequential decision-making problem and propose a solution using the RL paradigm \cite{xie2021personalized}.
Notably, these works prioritize the identification of optimal weights that can achieve Pareto efficiency across multiple objectives, rather than solely focusing on generating optimal results for specific preference weights. More recently, several studies~\cite{raychaudhuri2022controllable, navon2020learning, lin2020controllable} propose to learn the entire trade-off curve for MTL problems via hypernets, which has advantages in terms of runtime efficiency at the training phase for multiple preferences and real-time preference control at the inference phase. Inspired by the work of \cite{raychaudhuri2022controllable, lin2020controllable}, we propose to use hypernets to construct CMR.

\section{Preliminary}\label{sec:preliminary}
Given a set of $M$ candidate items $\mathcal{C}=\{c_i\}_{1\leq i \leq M }$, a user $u\in \mathcal{U}$, and a list reward function $R_{\boldsymbol{w}}(\cdot)$ parameterized by the preference weights $\boldsymbol{w}$, our goal is to find the optimal item list $L_{\boldsymbol{w}}^*$ among all possible lists composed by items from $\mathcal{C}$:
\begin{equation*}
L_{\boldsymbol{w}}^*=\underset{L}{\arg\max}~ R_{\boldsymbol{w}}(L(u, \mathcal{C})),
\end{equation*}
where each list is of length $N$ and it is obvious that $N\leq M$. 
In this paper, we assume that $R_{\boldsymbol{w}}$ takes a linear form:
\begin{equation*}
R_{\boldsymbol{w}}(L(u, \mathcal{C}))=\sum_{i=1}^{n_U}w_iU_i(L(u, \mathcal{C})),
\end{equation*}
where $U$ stands for utility for a certain objective and $n_U$ is the number of objectives of concern. Since $\boldsymbol{w}$ indicates the relative importance of utilities, it is called the preference weight. In previous multi-task learning works, $R_{\boldsymbol{w}}$ is fully predefined with a fixed $\boldsymbol{w}$ and a model parameterized by $\boldsymbol{\theta}$ is trained to approximate the mapping from $(u, \mathcal{C})$ to $L_{\boldsymbol{w}}^*$ for the specific $\boldsymbol{w}$. Although the calculation of the utility functions remains predefined, we treat the preference weight $\boldsymbol{w}$ as a variable, which can take any value from a feasible set, and we hope to find the optimal list $L_{\boldsymbol{w}}^*$ for any given $\boldsymbol{w}$. A brute force solution is to train a model for each feasible value of $\boldsymbol{w}$. It can be prohibitively time and resource consuming even for a small number of $\boldsymbol{w}$. 
One feasible solution is to enable one trained re-ranking model to serve all feasible values of $\boldsymbol{w}$, as will be shown in our CMR framework.

We represent each user $u$ as an embedding vector $\boldsymbol{x}_u \in \vmathbb{R}^{d_u}$. Similarly, a candidate item is represented by a feature vector $\boldsymbol{x} \in \vmathbb{R}^{d_i}$. A candidate set is represented as a bag of embeddings ${\{\boldsymbol{x}_i}\}_{i=1}^M$ and a list of $N$ items can be represented as a matrix $X \in \vmathbb{R}^{d_i\times N}$, which is a stacking of individual item embeddings $X = [\boldsymbol{x}_1, \boldsymbol{x}_2, \cdots, \boldsymbol{x}_N]$. A user-item engagement is represented by $\boldsymbol{y}\in \vmathbb{R}^{d_e}$ where each element in the engagement vector $\boldsymbol{y}$ could be, for example, exposure, click, purchase, or video viewing time, etc., depending on the specific applications. The user-list engagement is represented by the matrix $Y=[\boldsymbol{y}_1, \boldsymbol{y}_2, \cdots, \boldsymbol{y}_N] \in \vmathbb{R}^{d_e \times N}$. Note that we do not have the engagement of all $M$ candidate items, since only $N$ of them are in the final recommended list and a user does not have access to the rest.
An offline log sample is a tuple of $(u, \mathcal{C}, X, Y)$ and a log data set   $\mathcal{D}$  is $\{(u_i, \mathcal{C}_i, X_i, Y_i)\}_{i=1}^{n_D}$ where $n_D$ is the size of the data set. A utility function is a mapping from $(u, X)$ to a scalar, $U:\vmathbb{R}^{d_u} \times \vmathbb{R}^{d_i\times N} \rightarrow \vmathbb{R}$. Scalars $d_i$, $d_e$, and $d_u$ denote the respective embedding dimensions.

\section{Our Proposal}\label{sec:framework}
In the following sections, we will introduce the overall framework of CMR, our re-ranking model design, the business-oriented tasks, and the model training algorithm, respectively.
\subsection{The CMR Framework}
\begin{figure}[t]
    \centering
    \includegraphics[width=0.46\textwidth]{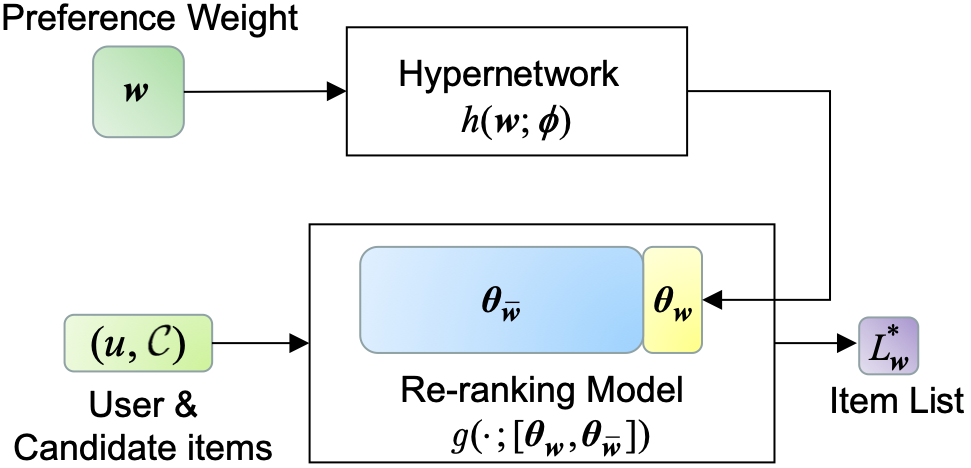}
    \caption{The proposed CMR framework.}
    \label{fig:cmr}
\end{figure}

Fig.~\ref{fig:cmr} shows the basic structure of the CMR framework, which consists of a hypernetwork and an arbitrary multi-objective re-ranking model. The hypernetwork $h(\boldsymbol{w}; \boldsymbol{\phi})$ takes in the preference weights $\boldsymbol{w}$ as inputs and generates parameters for the re-ranking model, which helps the re-ranking to generate recommendation lists according to the specific preference weights $\boldsymbol{w}$.
A re-ranking model based on modern deep neural networks can be as large as 50GB in size with tens of billions of parameters. It is infeasible for the hypernetwork to generate all of them. 
To circumvent this issue, one can split the parameter of the re-ranking model into $\boldsymbol{w}$-sensitive ones $\boldsymbol{\theta}_{\boldsymbol{w}}$ and $\boldsymbol{w}$-insensitive ones $\boldsymbol{\theta}_{\bar{\boldsymbol{w}}}$. 
The majority of $\boldsymbol{\theta}$ belongs to $\boldsymbol{\theta}_{\bar{\boldsymbol{w}}}$ which may include item embedding layers and representation learning layers. In contrast, $\boldsymbol{\theta}_{\boldsymbol{w}}$ can be a very compact part of $\boldsymbol{\theta}$, such as the last a few layers of a re-ranking model. 
Moreover, significant invasive structure modifications of a model usually require retraining of the entire model, which can be prohibitively expensive in industrial applications. By splitting $\boldsymbol{\theta}$, one can only modify the part involving $\boldsymbol{\theta}_{\boldsymbol{w}}$ and keep the $\boldsymbol{\theta}_{\bar{\boldsymbol{w}}}$ part intact, which helps to accommodate existing deployed models into the CMR framework.

Inspired by Conditional GAN~\cite{mirza2014conditional}, we use conditional training in the CMR framework as shown in Algorithm~\ref{tab:conditional_training}. For each training sample or batch, a preference weight $\boldsymbol{w}$ is sampled from its predefined feasible distribution $\mathcal{P}_{\boldsymbol{w}}$. Then it is fed into the hypernetwork to generate $\boldsymbol{\theta}_{\boldsymbol{w}}$ and the re-ranking model is invoked to generate a list $L$ with parameters $[\boldsymbol{\theta}_{\boldsymbol{w}}, \boldsymbol{\theta}_{\bar{\boldsymbol{w}}}]$. The reward function $R_{\boldsymbol{w}}$ evaluates $L$ with the sampled $\boldsymbol{w}$. Finally, the parameters of the hypernetwork $\boldsymbol{\phi}$ are updated according to the value of $R_{\boldsymbol{w}}$. Optionally $\boldsymbol{\theta}_{\bar{\boldsymbol{w}}}$ can be updated jointly if the re-ranking model is not fully trained. Various optimization methods can be used to update $\boldsymbol{\phi}$ and $\boldsymbol{\theta}_{\bar{\boldsymbol{w}}}$, from simple gradient descent to complex reinforcement learning algorithms. In principle, lists with high rewards should be generated with high probability. The feasible distribution of the preference weight $\mathcal{P}_{\boldsymbol{w}}$ can be derived from domain knowledge, business considerations, or practice experience. For example, relevance is typically considered more important than diversity, resulting in a higher weight for relevance utility than diversity utility. To test the generalization ability of CMR, we attempt to minimize our reliance on prior knowledge in the experiments and simply sample the preference weight of each utility function from a uniform distribution $U(0,w_i^\text{max})$ independently, where $w_i^\text{max}$ is the maximum value for a given preference weight, ranging from 0 to 1. The hypernetwork is deployed along with the re-ranking model and $\boldsymbol{w}$ is specified in real-time for each case during serving. The CMR framework enables a re-ranking model to generate the optimal recommendation list for any given $\boldsymbol{w}$. 
Determining optimal $\boldsymbol{w}$ at serving time is beyond the scope of this paper.

\subsection{The Proposed Re-ranking Model}
Our re-ranking model is based on the AE framework which can incorporate various utilities easily. It contains two main modules: an actor for list generation and an evaluator for list evaluation.
\subsubsection{The Actor Module}\label{sec:actor}
\begin{figure*}[h]
    \centering
    \setlength{\abovecaptionskip}{0.05cm}
    \includegraphics[width=0.8\textwidth]{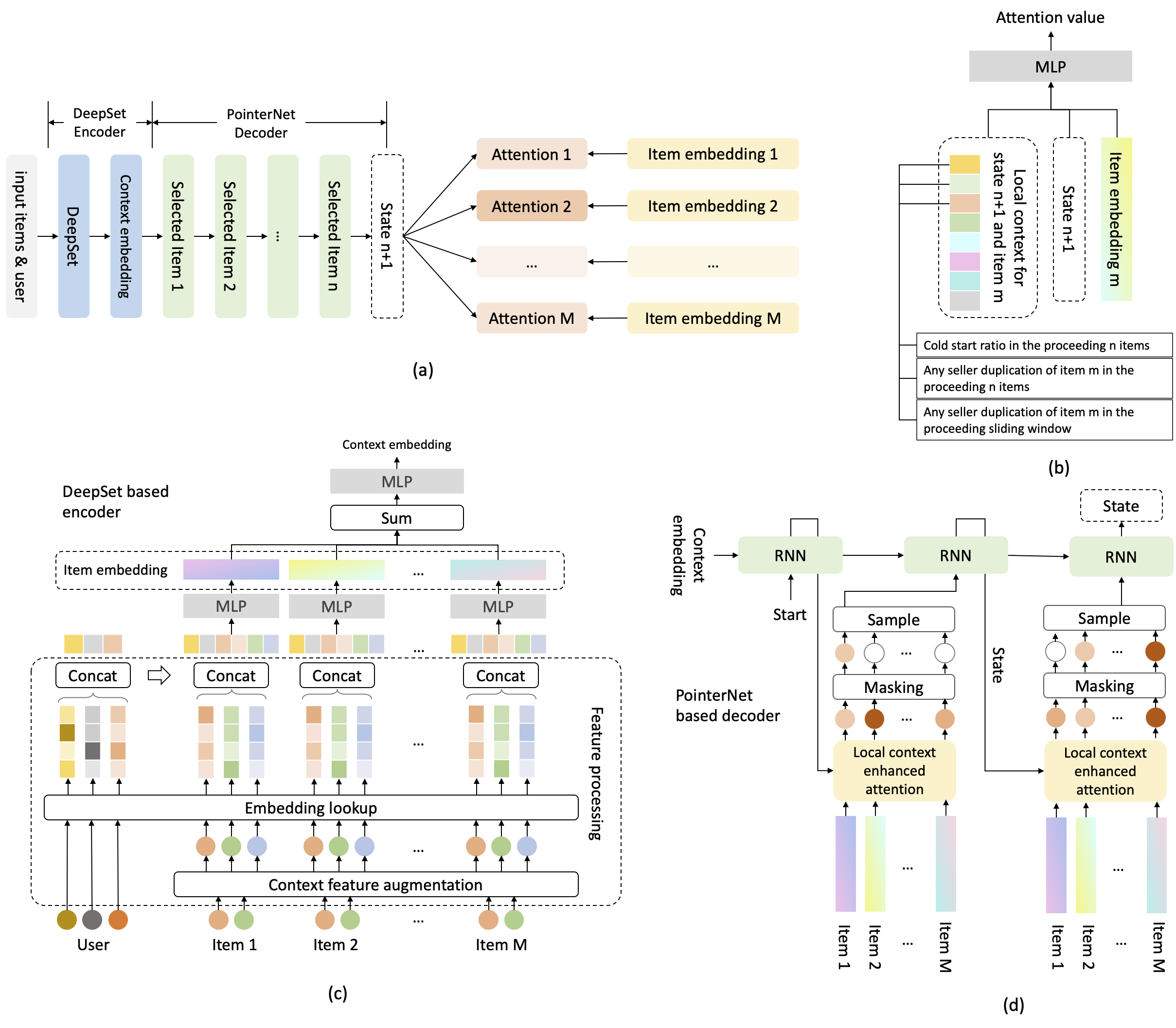}
    
    \caption{Model structure of the Actor. (a) the encoder-decoder structure; (b)  local context enhanced attention for item selection; (c)~the DeepSet-based encoder; (d) the PointerNet-based decoder.}
    \label{fig:actor}
    
\end{figure*}

The Actor has an encoder-decoder structure as illustrated in Fig.\ref{fig:actor}(a).
Before feeding into the encoder, feature augmentation and embedding lookup are carried out. The augmented features of an item come from the candidate set, such as the ranking of the item in terms of historical Click Through Rate (CTR) within the set. It is a very efficient and simple way of informing a model of the relationship between an individual item and its containing candidate set. Then ID features are transferred to real-valued embeddings to facilitate numerical computation.

\begin{algorithm}[t]
    \caption{Conditional training in the CMR framework}
    \label{tab:conditional_training}
    \begin{algorithmic}[1]
    \REQUIRE hypernetwork $h(\cdot; \boldsymbol{\phi})$ parameterized by $\boldsymbol{\phi}$; re-ranking model $g(\cdot;[\boldsymbol{\theta}_{\boldsymbol{w}}, \boldsymbol{\theta}_{\bar{\boldsymbol{w}}}])$ parameterized by the $\boldsymbol{w}$-sensitive and insensitive parameters; a list reward function $R_{\boldsymbol{w}}$; 
    training dataset $\mathcal{D}=\{(u, \mathcal{C}, X, Y)_i\}_{i=1}^{n_D}$; a feasible distribution of the preference weight $\mathcal{P}_{\boldsymbol{w}}$ 
    \ENSURE a learned hypernetwork $h(\cdot; \boldsymbol{\phi}^*$) and a re-ranking model $g$ with insensitive parameters $\boldsymbol{\theta}_{\bar{\boldsymbol{w}}}^*$

    \REPEAT
      \STATE sample a training instance $(u, \mathcal{C}, X, Y)\sim \mathcal{D}$
      \STATE sample a preference weight $\boldsymbol{w}\sim\mathcal{P}_{\boldsymbol{w}}$
      \STATE $\boldsymbol{\theta}_{\boldsymbol{w}}\leftarrow h(\boldsymbol{w}; \boldsymbol{\phi})$\COMMENT{run the hypernetwork}
      \STATE $L\leftarrow g((u, \mathcal{C}); [\boldsymbol{\theta}_{\boldsymbol{w}}, \boldsymbol{\theta}_{\bar{\boldsymbol{w}}}])$ \COMMENT{run the re-ranking model}
      \STATE evaluate the list by $R_{\boldsymbol{w}}(L)$ 
      \STATE update $\boldsymbol{\phi}$ to modify the generating probability of $L$ according to $R_{\boldsymbol{w}}(L)$ with  proper algorithms
      \STATE optionally, update $\boldsymbol{\theta}_{\bar{\boldsymbol{w}}}$ to modify the generating probability of $L$ according to $R_{\boldsymbol{w}}(L)$ with  proper algorithms if the re-ranking model has not been well trained
    \UNTIL{Converge}
    \RETURN $\boldsymbol{\phi}^*$ and $\boldsymbol{\theta}_{\bar{\boldsymbol{w}}}^*$
    \end{algorithmic}
\end{algorithm}

\textbf{The DeepSet-based Encoder.}
DeepSet~\cite{zaheer2017deep} is selected as the encoder because it is insensitive to the order of the input items, as shown in Fig.\ref{fig:actor}(c). It is preferable because, in complex applications, it is not trivial to find a good initial list mixing texts, images, and videos. Poor initial ordering may hurt the performance of a re-ranking model. It is neither necessary to provide an initial list since $(u, \mathcal{C})$ contains all information a re-ranking model needs to know to generate a good list. The context embedding $\boldsymbol{e}_c$ is calculated as
\begin{equation*}
\boldsymbol{e}_c=MLP_2 \left(\sum_{i=1}^MMLP_1([\boldsymbol{x}_i; \boldsymbol{x}_u]) \right),
\end{equation*}
where $\boldsymbol{x}_u$ means user vector and $[\boldsymbol{x}_i; \boldsymbol{x}_u]$ means the concatenation of the two vectors. $MLP$ stands for multilayer perceptron. 
Note that the output of $MLP_1$ is used as the item embedding $\boldsymbol{e}$ for each item. 

\textbf{The PointerNet-based Decoder.}
The decoder is based on PointerNet~\cite{vinyals2015pointer} which picks one item from the candidate set at one time, updates context information immediately, and then picks the next, as shown in Fig.\ref{fig:actor}(d). Item selection is based on a local context enhanced attention mechanism as shown in Fig.\ref{fig:actor}(b). 

At the very beginning of decoding, the context embedding $\boldsymbol{e}_c$ is used as the initial hidden state of a Recurrent Neural Network (RNN) cell and a special token "start" is fed into the RNN cell as the input. After that, at each step, the output embedding of the RNN cell acts as the state embedding $\boldsymbol{e}_s$, which is supposed to contain all necessary information for item selection. The local context enhanced attention $\boldsymbol{a}\in \vmathbb{R}^M$ is calculated as 
\begin{equation}
\label{eq:local_attention}
\boldsymbol{a}=\operatorname{softmax}([MLP_3([\boldsymbol{e}_s; \boldsymbol{e}_i; \boldsymbol{e}_{si}])]_{i=1}^M),
\end{equation}
where $\boldsymbol{e}_{i}$ is the item embedding and
 $\boldsymbol{e}_{si}$ is the local context embedding. It is termed ``local'' because the context is specific to the current state and the current candidate item. As illustrated in Fig.~\ref{fig:actor}(b), the introduction of the local context is to facilitate the model's understanding of business rules and prior knowledge of a good list. For example, one may want an item list from diverse sellers. It can be achieved by controlling seller duplication in the result list. If a candidate item introduces a seller duplication, its attention value should be smaller so that it has a lower chance to be selected. Although the piece of information regarding seller duplication can potentially be abstracted implicitly by a complex neural network, it is not necessary and is low efficient. Instead, we can directly add a feature ``the number of seller duplication of the current item in the preceding list'' into the local context embedding $\boldsymbol{e}_{si}$. 

Masking is applied to the attention values for two main reasons. First, an item that has been selected in the preceding steps should not be selected again. So their attention values are masked to zero. Second, business rules may ask a model to place a certain item at a certain position in the result list. In this case, the attention values of all items other than the target one are masked to zero. Incorporating a sampling mechanism is crucial for an actor because it lets the actor try different actions and ultimately find the optimized list-generating policy. We use Thompson Sampling, which samples an item proportionally to its masked attention value.

\subsubsection{The Evaluator Module}\label{sec:evaluator}
The evaluator contains an ensemble of utility functions. Recall that a utility function is a mapping from $(u, X)$ to a scalar, $U:\vmathbb{R}^{d_u} \times \vmathbb{R}^{d_i\times N} \rightarrow \vmathbb{R}$, indicating the quality of the list in one perspective.  Some of the utility functions come from business-oriented requirements and prior knowledge of good lists, such as item diversity. These utilities have analytical formulas and can be calculated directly. Other utilities, such as the probability of user-list engagement, have to be predicted by models, which are trained with offline data. Once trained, they are used as predefined utility functions. This section introduces the model used for the predictions.

\begin{figure*}[h]
    \centering
    \setlength{\abovecaptionskip}{0.05cm}
    \includegraphics[width=0.85\textwidth]{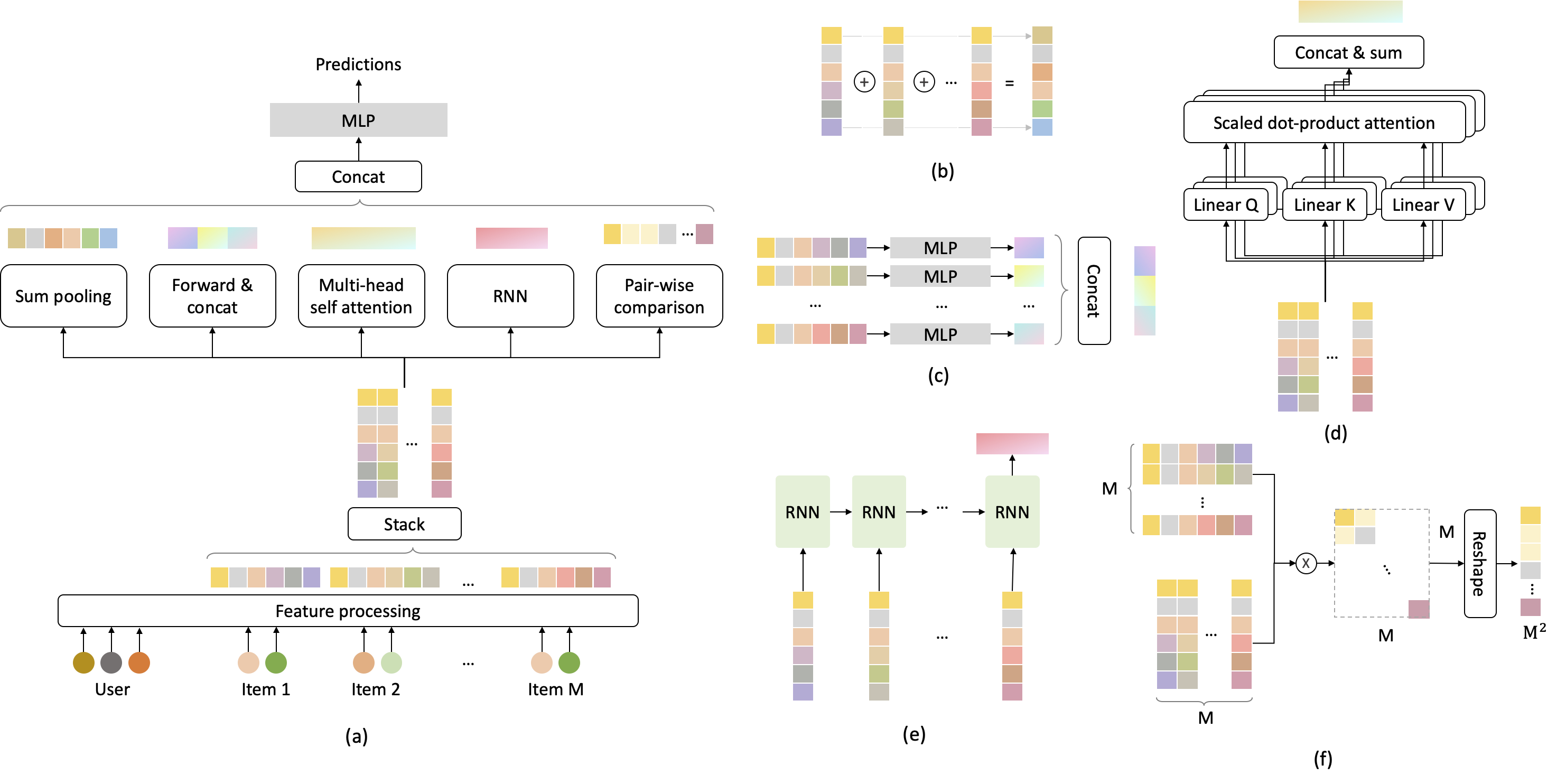}
    
    \caption{Model structure of the Evaluator. (a) the overall structure; (b) the sum pooling channel; (c) the forward \& concat channel; (d) the multi-head self-attention channel; (e) the RNN channel; (f) the pair-wise comparison channel. }
    \label{fig:evaluator}
    
\end{figure*}

The overall structure of the evaluator model is shown in Fig.\ref{fig:evaluator}(a). After the same feature processing as that of in the actor, the item list is represented by a matrix $E=[\boldsymbol{e}_1, \boldsymbol{e}_2, ..., \boldsymbol{e}_N]$. It is processed by 5 channels that focus on different aspects of the list. The results are then concatenated and transferred to the final predictions through a $MLP$. The first channel is "sum pooling", which is an element-wise sum of the item embeddings as shown in Fig.\ref{fig:evaluator}(b), resulting in 
$
\boldsymbol{e}_{sp} := \sum_{i=1}^N\boldsymbol{e}_i,
$
The second channel is "forward \& concat" which is good at abstracting useful information from each item as in Fig.\ref{fig:evaluator}(c) and results in an embedding $\boldsymbol{e}_{fc}$ as 
\begin{equation}
\label{eq:forward_concat}
\boldsymbol{e}_{fc}=[MLP_4(\boldsymbol{e}_1);MLP_4(\boldsymbol{e}_2); \cdots ;MLP_4(\boldsymbol{e}_N)].
\end{equation}
The third channel is "multi-head self-attention" as in Fig.\ref{fig:evaluator}(d) which aims to capture mutual influence between items. It outputs $\boldsymbol{e}_{mh}$ as 
\begin{equation*}
\boldsymbol{e}_{mh}=reduce\_sum([head_h(W_{h}^qE, W_{h}^kE, W_{h}^vE )]_{h=1}^H), \\
\end{equation*}
where $W^q$, $W^k$, and $W^v$ are linear projection matrices and $H$ is the number of heads. $head$ stands for scaled dot-product attention. %
The $reduce\_sum$ of a matrix $X=[\boldsymbol{x}_1, \boldsymbol{x}_2, \ldots, \boldsymbol{x}_n]$ means $
reduce\_sum(X) = \sum_{i=1}^n\boldsymbol{x}_i.
$
The fourth channel is "RNN" as in Fig.\ref{fig:evaluator}(e)  which reveals the evolution trend of the list. The result embedding $\boldsymbol{e}_{rnn}$ is the output of the RNN cell at the final step. 
The fifth channel is "pair-wise comparison" as in Fig.\ref{fig:evaluator}(f). As suggested by its name, it compares each item pair using inner product and the result embedding $\boldsymbol{e}_{pc}=[\langle\boldsymbol{e}_i, \boldsymbol{e}_j \rangle ]_{1\leq i \leq N, i < j \leq N}.
$
The final $Predictions$ are obtained via a $MLP$
\begin{equation*}
\label{eq:evaluator_prediction}
Predictions = MLP_5([\boldsymbol{e}_{sp};\boldsymbol{e}_{fc}; \boldsymbol{e}_{mh}; \boldsymbol{e}_{rnn}; \boldsymbol{e}_{pc}]).
\end{equation*}

\subsection{Business-Oriented Utilities}\label{sec:business_oriented_utilities}
We classify business-oriented tasks into four main categories: fixed-position insertions, flow control, diversity, and group ordering. Many real-life complex business problems can be broken down into these tasks or their combinations. Traditionally, these tasks are handled separately in a pipeline manner, leading to sub-optimal performance. This section incorporates all the tasks into our proposed re-ranking model which provides an end-to-end optimal solution. 

\subsubsection{Flow Control}\label{sec:flow_control}
Flow control ensures that the contents from certain groups get enough exposure to customers. It comes from several perspectives: from the fairness perspective where the groups are protected minorities, and from a platform ecology perspective where contents by new providers are distributed with priority.

The effectiveness of flow control is measured by exposure ratio. 
We may want the exposure ratio of a certain group to be greater than a predefined threshold. The flow control utility is defined as
\begin{equation}
\label{eq:flow_control_u1}
U_{g}^{f} = -\frac{1}{|b|}\sum_{p \in b}\vmathbb{I}\left( \frac{\sum_{i \in p}\vmathbb{I}\left( i \in \mathcal{I}_{g} \right)}{|p|} \le t_{g}^{e} \right),
\end{equation}
where the superscript $f$ means flow control, $p$ means a page, $b$ means a training batch, and $t^e$ means the exposure ratio threshold. 
With a little abuse of notations, $i$ stands for the item at position $i$ as well as the position index of an item in the list, as long as it does not cause confusion. $\mathcal{I}_{g}$ is a set containing all items belonging to the group $g$. $|p|$ stands for the number of slots in the page. $\vmathbb{I}$ is the indicator function. Please note the negative sign at the beginning of the right-hand side of the equation. It means recommendation lists violating the requirement of flow control will receive a penalty. 

Eq.~\eqref{eq:flow_control_u1} is too strict since it requires all pages to satisfy the threshold. We only want to control the exposure ratio at the population level, rather than on each page. Although the exposure ratio of the whole population is not accessible during training, the exposure ratio of a batch can work as a handy approximation:
\begin{multline}
\label{eq:flow_control_u2}
U_{g}^{f} = -\frac{1}{|b|}\vmathbb{I}\left( \text{ RatioInBatch}(g)\le t_{g}^{e}\right)\sum_{p \in b}\vmathbb{I} \left( \frac{\sum_{i \in p}\vmathbb{I}( i \in \mathcal{I}_{g} )}{|p|} \le t_{g}^{e} \right).
\end{multline}
where $\text{ RatioInBatch}(g) =  \frac{\sum_{p \in b}\sum_{i \in p}\vmathbb{I}( i \in \mathcal{I}_{g} )}{|b||p|}$ is a gating. As long as the exposure ratio of the whole population meets the requirement, the exposure ratio of individual pages is no longer a concern.

In the scenarios where customers perceive items in a result list one by one, customers may leave at any time, and items at the bottom of a list may not be actually seen by the customers, which makes the above exposure ratio calculation inaccurate. To handle it, we can add another position term
\begin{multline}
\label{eq:flow_control_u3}
U_{g}^{f} =  -\frac{1}{|b|}\vmathbb{I}\left( \text{ RatioInBatch}(g)\le t_{g}^{e}\right)\sum_{p \in b}\vmathbb{I} \left( \frac{\sum_{i \in p}\vmathbb{I}( i \in \mathcal{I}_{g} )}{|p|} \le t_{g}^{e} \right)- \\
\frac{1}{|b|}\vmathbb{I}\left( \text{ PosInBatch}(g) \ge t_{g}^{p} \right) 
\sum_{p \in b}\vmathbb{I}\left( \frac{\sum_{i \in p}\text{pos}_{i}\vmathbb{I}( i \in \mathcal{I}_{g} )}{\sum_{i \in p}\vmathbb{I}( i \in \mathcal{I}_{g} )} \ge t_{g}^{p} \right),
\end{multline}
where  
$
\text{ PosInBatch}(g) =  \frac{\sum_{p \in b}\sum_{i \in p}\text{pos}_{i}\vmathbb{I}( i \in \mathcal{I}_{g} )}{\sum_{p \in b}\sum_{i \in p}\vmathbb{I}( i \in \mathcal{I}_{g} )} ,
$
and $pos_i$ means the exposure position of item $i$ and $t^p$ is the exposure position threshold. Note that Eq.~\eqref{eq:flow_control_u1}, Eq.~\eqref{eq:flow_control_u2}, and Eq.~\eqref{eq:flow_control_u3} show possible choices of flow control utility for a particular group $g$. Flow control for more than one group may co-exist in one model.

\subsubsection{Diversity}\label{sec:diversity}
Diversity means items in result lists should come from different groups. The groups may refer to sellers, categories, supply sources, as well as item exhibition types such as text, image, and video. The diversity utility of a group $g$ can be written as 
\begin{equation*}
U_{g}^{d} = \frac{1}{|b||p|}\sum_{p \in b}\sum_{i \in p} \vmathbb{I}(g_{i}\notin \mathcal{G}_{i-1}),
\end{equation*}
where the superscript $d$ means diversity and $g_i$ means the group of the item at position $i$. $\mathcal{G}_{i-1}$ is the group set of items in a sliding window before item $i$: $\mathcal{G}_{i-1}=\{g_{j}|j \in \{i-n_g^d,i-(n_g^d -1),\ldots,i-1\} ,j>0\}$ where $n_g^d$ is the length of the diversity window. Setting $n_g^d=\infty$ promotes whole list-wise diversity. 

\subsubsection{Group Ordering}\label{sec:group_ordering}
Group ordering means the items from certain groups are ranked before others with high probability. For example, in a new arrival promoting scene, it is preferred to show new items released within 3 days first and then others. Group ordering can be achieved by checking the group priority of item pairs
\begin{equation*}
U_{g}^{o} = \frac{1}{|b|}\sum_{p \in b}
\frac{1}{2|p||p-1|}\sum_{i \in p}\sum_{j>i}\vmathbb{I}(\text{priority}(g_{i}) \ge \text{priority}(g_{j})),
\end{equation*}
where the superscript $o$ means group ordering and $\text{priority}$ indicates the predefined priority function of groups. 

\subsubsection{Fixed Position Insertion}\label{sec:fixed_position_insertion}
Fixed position insertion is a very strict task, which means a given item must be placed at a specified position of a result list with 100\% probability. A customer may come to a recommendation scene by clicking a trigger item, it is mandatory to place the trigger item at the very top of a recommendation list. 
It may seem trivial at first glimpse since one can always add the trigger item to the top of a list after the list has been created. 
However, such a method may fail to handle the relations between multiple objectives simultaneously.
For example, adding a trigger item on top of a separately generated list may result in seller duplication of the first two items, which breaks the diversity requirement. Adding a utility function does not work well for this task because a utility cannot offer a strong probability guarantee. Instead, we use masking to achieve fixed position insertion as introduced in \ref{sec:actor}. 

\subsection{Model Training}\label{sec:model_training}
 The evaluator is fully trained before the actor in a purely supervised learning fashion. So far we only train classification models for the evaluator with the classic Cross Entropy:
$
\mathcal{L}_{eval} = -\sum_{i=1} y_i \log(p_i),
$ 
where $y_i \in \{0, 1\}$ is the one-hot label for a class and $p_i$ is the model prediction for the class. The models may predict whether the generated list will lead to a certain kind of user engagement, e.g. click, and the intensity of the engagement.

We use a REINFORCE-based method for actor training. In reinforcement learning, a policy is a mapping from a state to a probability distribution over possible actions which maximizes the total reward. In simple words, policy tells us which action to take in a given state. In our case, Eq.~\eqref{eq:local_attention} plays the essential role of a policy, where the state is $[\boldsymbol{e}_s; \boldsymbol{e}_i; \boldsymbol{e}_{si}]_{i=1}^M$ and the probability distribution over M possible items is $\boldsymbol{a}$. 
Suppose for a user $u$ and a candidate item set $\mathcal{C}=\{c_1, c_2, ..., c_M\}$, the actor generate a list $L(u, \mathcal{C})$ as $[c_{\pi_1}, c_{\pi_2}, ..., c_{\pi_N}]$. $\boldsymbol{\pi}$ is the indicator of a list, where $\pi_n=m$ means the $m$th item in the candidate set is placed in the $n$th position in the output list. Here we assume an arbitrary ranking in the candidate set to get the index. The loss function for the actor  $\mathcal{L}_{actor}$ is 
\begin{equation*}
\mathcal{L}_{actor} = -\left[R_{\boldsymbol{w}}(L(u, \mathcal{C})) - R_{\boldsymbol{w}}(L_{exp})\right]\sum_{n=1}^N \log(a_{\pi_n}),
\end{equation*}
where $L(u, \mathcal{C})$ is the list generated by the actor, $L_{exp}$ is the recorded exposure list, the reward difference between the two $R_{\boldsymbol{w}}(L(u, \mathcal{C})) - R_{\boldsymbol{w}}(L_{exp})$ severs as the advantage function. In this loss, we assume each action contributes equally to the advantage value. We recognize advanced training options such as Proximal Policy Optimization (PPO) \cite{schulman2017proximal}. Nevertheless, the REINFORCE-based method works well so we leave the choice of the options to further studies. We observe that gradient clipping greatly benefits the training. 
\renewcommand{\thefootnote}{\arabic{footnote}} 
\section{Experiments}\label{sec:experiment}
We conducted offline and online experiments to verify the effectiveness of the proposed CMR framework. The source code
of offline experiments has been share at~\url{https://github.com/lyingCS/Controllable-Multi-Objective-Reranking}. 

\subsection{Offline Experiments}\label{sec:offline}
\begin{table*}[htbp]
\caption{Offline evaluation results of sequential re-ranking models. The experiments are repeated 3 times with different random seeds. We display the mean performance and standard deviation. The best results of diversity and accuracy metrics under the condition that $\lambda$(acc\_prefer) is 0 and 1 are highlighted in bold respectively and the second-best results are underlined.}
\label{Controllable Experiment 1}
\small{
\begin{tabular}{l|l|lllllll}
\hline

			Method & $\lambda$ & map@5 & map@10 & ndcg@5 & ndcg@10 & ilad@5 & err\_ia@5 & err\_ia@10 \\ \hline
			Initial & \textendash & 0.59961 & 0.60321 & 0.68092 & 0.69531 &  0.64535 & 1.29619 & 1.32672 \\ \hline
			&  0 & 0.5924$\pm$0.0011 & 0.5963$\pm$0.0011 & 0.6750$\pm$0.0010 & 0.6901$\pm$0.0008 & \underline{0.6449$\pm$0.0005} & \underline{1.2957$\pm$0.0003} & \underline{1.3264$\pm$0.0002} \\
			& 0.5 & 0.6001$\pm$0.0012 & 0.6038$\pm$0.0013 & 0.6815$\pm$0.0010 & 0.6958$\pm$0.0010 & 0.6448$\pm$0.0002 & 1.2955$\pm$0.0001 & 1.3262$\pm$0.0001\\
			\multirow{-3}{*}{Seq2Slate} &  1 & \textbf{0.6021$\pm$0.0005} & \textbf{0.6057$\pm$0.0005} & \textbf{0.6829$\pm$0.0003} & \textbf{0.6972$\pm$0.0004} &  0.6444$\pm$0.0002 & 1.2954$\pm$0.0002 & 1.3261$\pm$0.0001\\ \hline
			& 0 & 0.5988$\pm$0.0007 & 0.6026$\pm$0.0007 & 0.6807$\pm$0.0006 & 0.6948$\pm$0.0005 & 0.6444$\pm$0.0005 & 1.2952$\pm$0.0005 & 1.3259$\pm$0.0004 \\
			& 0.5 & 0.6000$\pm$0.0005 & 0.6038$\pm$0.0004 & 0.6815$\pm$0.0003 & 0.6957$\pm$0.0003 & 0.6434$\pm$0.0002 & 1.2944$\pm$0.0001 & 1.3253$\pm$0.0000  \\
			\multirow{-3}{*}{EG-Rerank} & 1 & 0.6009$\pm$0.0012 & 0.6046$\pm$0.0011 & 0.6822$\pm$0.0009 & 0.6963$\pm$0.0009 &  0.6424$\pm$0.0007 & 1.2935$\pm$0.0005 & 1.3247$\pm$0.0003 \\ \hline
			& 0 & 0.5949$\pm$0.0046 & 0.5987$\pm$0.0045 & 0.6772$\pm$0.0036 & 0.6919$\pm$0.0034 & \textbf{0.6485$\pm$0.0004} & \textbf{1.2991$\pm$0.0003} & \textbf{1.3289$\pm$0.0004} \\
			& 0.5 & 0.5971$\pm$0.0035 & 0.6009$\pm$0.0034 & 0.6789$\pm$0.0028 & 0.6935$\pm$0.0026 & 0.6484$\pm$0.0007 & 1.2987$\pm$0.0001 & 1.3286$\pm$0.0002\\
			\multirow{-3}{*}{CMR~(ours)} & 1 & \underline{0.6016$\pm$0.0018} & \underline{0.6053$\pm$0.0017} & \underline{0.6825$\pm$0.0014} & \underline{0.6969$\pm$0.0013} &  0.6480$\pm$0.0008 & 1.2983$\pm$0.0003 & 1.3282$\pm$0.0001 \\ \hline

\end{tabular}
}
\end{table*}

\begin{table*}[htbp]
\caption{Offline comparisons of CMR and rule-based baselines. 
No standard deviation is reported since 
random seeds do not affect APDR and MMR experimental results. The best results of diversity and accuracy metrics under the condition that $\lambda$(acc\_prefer) is 0 and 1 are highlighted in bold respectively and the second-best results are underlined.}
\label{Controllable Experiment 2}
\small{
\begin{tabular}{l|l|lllllll}
\hline
Method & $\lambda$ & map@5 & map@10 & ndcg@5 & ndcg@10 & ilad@5 & err\_ia@5 & err\_ia@10 \\ \hline
Initial & \textendash &  0.5996 &  0.6032 &  0.6809 &  0.6953 &  0.6454 &  1.2962 &  1.3267 \\ \hline
& 0 & 0.5965 & 0.6001 & 0.6787 & 0.6932 & \textbf{0.6673} & \textbf{1.3153} & \textbf{1.3409}\\
& 0.5 & 0.5967 & 0.6005 & 0.6788 & 0.6933 & 0.6617 & 1.3101 & 1.3373\\
\multirow{-3}{*}{APDR} & 1 & 0.5971 & 0.6010 & 0.6789 & 0.6936 & 0.6453 & 1.2962 & 1.3268\\ \hline
 &  0 &  0.5992 &  0.6027 &  0.6808 &  0.6951 &  \underline{0.6656} &  \underline{1.3147} &  \underline{1.3402}\\
 &  0.5 &  0.5996 &  0.6032 &  0.6810 &  0.6953 & 0.6565 &  1.3054 &  1.3335\\
\multirow{-3}{*}{MMR} &  1 &  \underline{0.5996} &  \underline{0.6032} &  \underline{0.6809} &  \underline{0.6953} &   0.6454 &  1.2962 &  1.3267\\ \hline
			& 0 & 0.5949 & 0.5987 & 0.6772 & 0.6919 & 0.6485 & 1.2991 & 1.3289 \\
			& 0.5 & 0.5971 & 0.6009 & 0.6789 & 0.6935 & 0.6484 & 1.2987 & 1.3286\\
			\multirow{-3}{*}{CMR~(ours)} & 1 & \textbf{0.6016} & \textbf{0.6053} & \textbf{0.6825} & \textbf{0.6969} &  0.6480 & 1.2983 & 1.3282 \\ \hline

\end{tabular}}
\end{table*}
We trained CMR on the benchmark LibRerank\footnote{https://github.com/LibRerank-Community/LibRerank\label{LibRerank}} with the public recommendation dataset Ad2\footnote{https://tianchi.aliyun.com/dataset/56}. We conducted experiments to answer the following two questions: i) Can CMR take effect on various re-ranking models? ii) How does our method perform compared with other rule-based controllable re-ranking baselines?
\subsubsection{Experiment Setting}\label{sec:experiment setting}
We conducted our offline experiments on the public benchmark LibRerank\textsuperscript{\ref{LibRerank}}, which can automatically perform re-ranking experiments and integrate a major collection of re-ranking algorithms.

To answer the first question, we extend the single accuracy objective in the original benchmark to multi-objective by adding diversity objective. Specifically, inspired by MDP-DIV\cite{xia2017adapting}, we define the diversity reward function as the promotion of diversity metric caused by choosing the $i$-th item:
\begin{equation*}
R^{div}_i(L(u, \mathcal{C}))={ERR\_IA}[L_i(u, \mathcal{C})]-{ERR\_IA}[L_{i-1}(u, \mathcal{C})].
\end{equation*}
The loss function is designed to be
$
\mathcal{L}_{actor} := \lambda\mathcal{L}^{acc}_{actor} + (1-\lambda)\mathcal{L}^{div}_{actor},
$ 
where $\lambda\in[0,1]$ is the trade-off parameter 
and $\mathcal{L}^{acc}_{actor},\\ \mathcal{L}^{div}_{actor}$ stand for losses in terms of utilities of accuracy and diversity. 

\textbf{Dataset:}
The original Ad dataset records 1 million users and 26 million ad display/click logs, with 8 user profiles (e.g., id, age, and occupation), 6 item features (e.g., id, campaign, and brand). LibRerank transformed the records of each user into ranking lists according to the timestamp of the user browsing the advertisement. Items that have been interacted with within 5 minutes are sliced into a list. 
The final Ad dataset contains 349,404 items and 483,049 lists. 
\textbf{Baselines:}
We chose several representative methods as baselines:
\textbf{Seq2Slate}~\cite{bello2018seq2slate}: a sequence-to-sequence model that formulates the re-ranking as a sequence generation problem, and sequentially selects the next items by pointer network;
\textbf{EG-Rerank}~\cite{huzhang2021aliexpress}: a re-rank model that adopts an evaluator generator paradigm—with a generator to generate feasible permutations and an evaluator to evaluate the listwise utility of each permutation;
\textbf{MMR}~\cite{carbonell1998use}: a heuristic approach with the documents selected sequentially according to maximal marginal relevance;
\textbf{APDR}~\cite{teo2016adaptive}: a learning model that aims to improve the diversity and personalization of image search results.  

\textbf{Evaluation metrics:}
We selected MAP, NDCG as the accuracy metrics, and ILAD\cite{hao2021re}, ERR\_IA\cite{yan2021diversification} as the diversity metrics. Specifically, we chose MAP@5, MAP@10, NDCG@5, NDCG@10, ILAD@5, ERR\_IA@5 and ERR\_IA@10. 
Note that MAP and NDCG are greedy metrics, which means the highest evaluation is reached when items are ranked strictly according to their "relevance". However, such lists may not lead to the best online performance. Instead, the AE framework-based re-ranking models aim to maximize list-wise user engagement, without putting too much emphasis on in-page item ordering. The user engagement is predicted by the evaluator. As a result, the best online performing  AE re-ranking models may seem sub-optimal in terms of MAP and NDCG. We stick with the metrics because they are popular and can be easily applied to various methods, whether based on the AE framework or not.

\subsubsection{Performance Analysis}\label{sec:performance analysis}

To answer the first question, we investigate the following models in the CMR framework: Seq2Slate, EG-Rerank, and ours. The results are reported in Table ~\ref{Controllable Experiment 1}. The first row of the table shows the metrics of the list after sorting by ranker. In this experiment, the ranker we used is lambdaMART\cite{burges2010ranknet}. Then we tested the two baselines and our model at three different accuracy preference(0, 0.5, 1), and counted their 4 accuracy metrics and 3 diversity metrics. Note that any permutation makes no difference to the calculation of ILAD@10, so we do not present this metric in the table. We notice that as the accuracy preference increases, the accuracy metrics show an upward trend, and the diversity metrics are basically the opposite. 
\begin{figure}[t]
    \centering
    \includegraphics[width=0.48\textwidth]{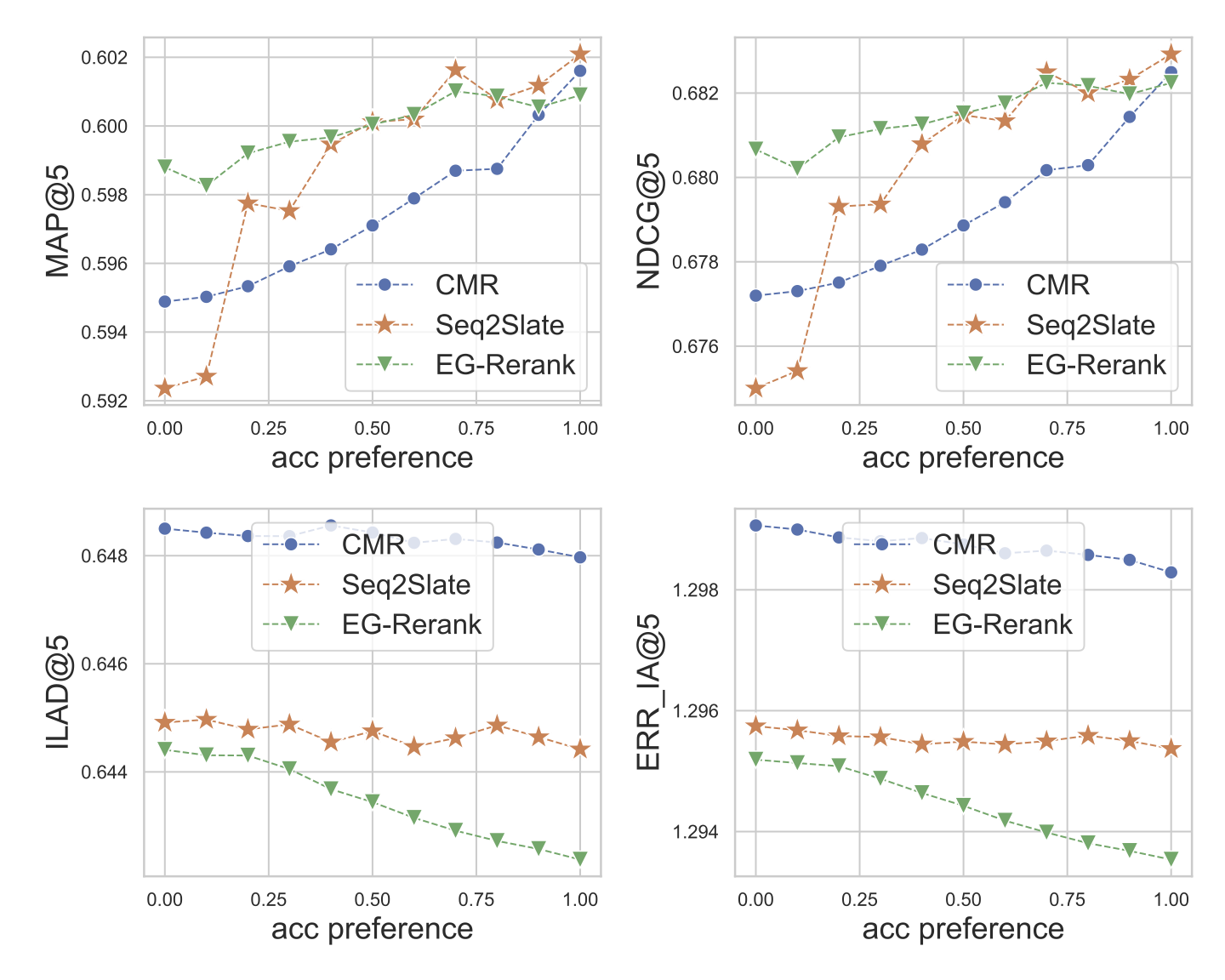}
    \caption{\mbox{Controllable effects of sequential re-ranking models.}}
    \label{fig:controllable evaluation}
\end{figure}
We then conducted controllable experiments of the above three re-ranking models on the test set. The final results are shown in Fig.~\ref{fig:controllable evaluation}, where the horizontal axis is the preference for accuracy and the vertical axis is the corresponding metric. The upper two subgraphs are the accuracy metrics MAP@5 and NDCG@5, and the lower ones are the diversity metrics ILAD@5 and ERR\_IA@5. As we can see, there is a clear trend that as the accuracy preference increases, the corresponding accuracy metrics increase and the corresponding diversity metrics decreases, with the most obvious and smoothest change in CMR. It can also be seen that for the accuracy metrics, the effect of Seq2slate is slightly better than that of the other two re-ranking models, followed by CMR. But for diversity metrics, CMR is much better. So our CMR framework can be well adapted to various re-ranking models.

For the second question, We compared our model with the existing rule-based controllable methods APDR and MMR. The experiment results are performed in Table~\ref{Controllable Experiment 2}. We observe that though under diversity metrics CMR underperforms APDR and MMR, it can break through the limitations of the rule-based methods and outperforms these baselines in terms of accuracy.

\subsection{Online Experiments}\label{sec:online}
We conducted online experiments on the "Subscribe" scene in the Taobao App, a leading e-commerce platform in China. Its main entrance is the "Subscribe" button on top of Taobao's main landing page, and it is a stream of various elements including item lists, posters, coupons, etc. 

\begin{figure}[!htbp]
    \centering
    \includegraphics[width=0.47\textwidth]{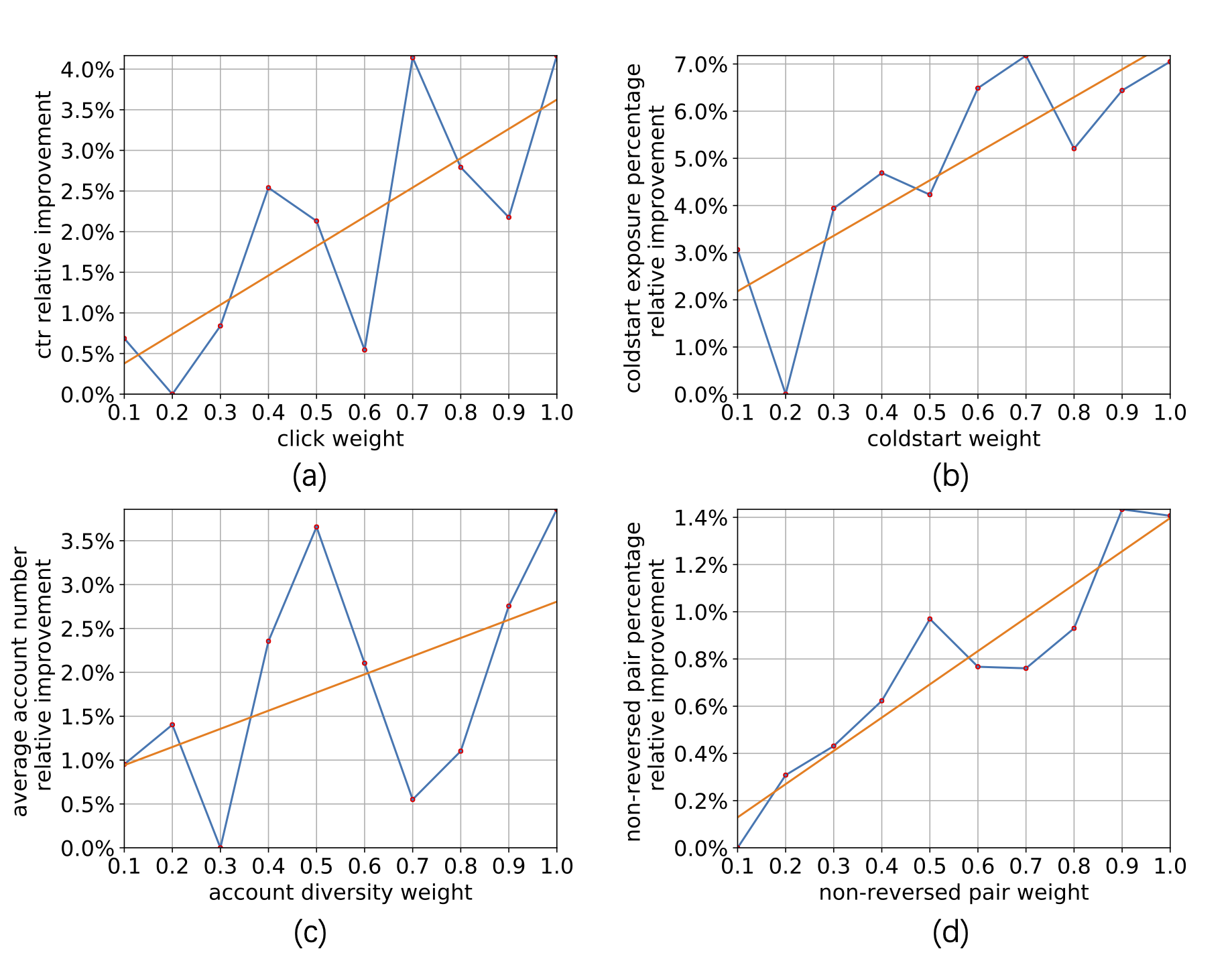}
    \caption{Online metric changes w.r.t. preference weights. The orange lines are the linear fitting of the blue metric curves. }
    \label{fig:online_1}
    
\end{figure}

In the first experiment, we evaluated whether an online performance metric changes well with a corresponding preference weight. We train a CMR re-ranking model from scratch and assign random preference weight during serving. We picked one representative from each of the four types of business-oriented tasks:  a  flow control utility to ensure the exposure ratio of cold start contents as in  Eq.~\eqref{eq:flow_control_u3}, a seller account diversity utility to promote the number of sellers in result lists,  a group ordering utility to place more new contents in the front of result lists, and a traditional click utility to promote user engagement. Trigger contents are guaranteed to sit at the top of result lists with fixed position insertion masking. The result is shown in Fig.~\ref{fig:online_1}, where the horizontal axis is the weight of a utility and the vertical axis is the relative improvement of a corresponding metric online. Logged samples are grouped into successive bins according to preference weights, and the average metric within each bin is calculated. The metrics are normalized so that the minimum value is 1 and the relative improvement is calculated with respect to the minimal value. As we can see, there is a clear trend that as a preference weight goes up, the corresponding online metric increases. Although the range of the preference weights is the same from 0 to 1, the adaptive ranges of the metrics vary from 1.4\% to 7\%. It is likely an essential reflection of the nature of the recommender system. Note that currently the CMR model takes 20 inputs and outputs a list of 10, which limits the adaptive ranges of the metrics. We are working on increasing the input size. The blue metric curves are not ideally monotonical. We mainly attribute it to data sparsity. As online experiments may hurt real users' experience, we only use a small portion of the App traffic.

In the second experiment, we aim to show the superiority of the joint optimal solution provided by the proposed re-ranking model. The baseline is a classic module pipeline solution. In the first module, the click probability of each content is predicted as the basic ranking score to form the initial recommend list. In the second module, a freshness bias is added to the ranking score so that new contents are placed relatively in front of result lists. In the third module, a heuristic diversity algorithm is applied to promote seller diversity\cite{teo2016adaptive}. In the fourth module, cold start contents are inserted into the list. In the final module, trigger contents are placed at the top of the list. This is the default solution of the online system and the involved hyper-parameters are kept tuned manually for years. Our CMR re-ranking model gets rid of the pipeline and offers an end-to-end one-step solution. One CMR model is trained and the preference weights are tuned manually online. The model config, such as the utilities involved, is the same as in the first online experiment. As no other re-ranking model considers such a big scope, we show the A/B test results of the base solution and the CMR solution only. Table~\ref{tab:online_ab} shows the results of a 7-day online A/B test. The first four metrics are directly related to the utility functions in the CMR model and all of them are improved, which are content click number per user, seller exposure number per user, cold start exposure ratio, and chronological ordering. An interesting observation is that stay time per user and content exposure number per user are also improved, although we do not model them directly. In our experience, it is a sweet byproduct of AE based re-ranking models. 
 \begin{table}[!htb]
\caption{Online A/B test results.}
\setlength{\belowcaptionskip}{0.05cm}
\label{tab:online_ab}
\small{
\begin{tabular}{lr}
\hline
Metric & Relative improvement \\
\hline
content click number per user & 0.62\% \\
seller exposure number per user & 2.43\% \\
cold start exposure ratio & 4.27\% \\
chronological ordering & 1.40\% \\
stay time per user & 1.41\% \\
content exposure number per user  & 0.73\% \\
\hline

\end{tabular}}
\end{table}

\section{Conclusion}
This paper presents a controllable multi-objective re-ranking (CMR) framework to adapt the recommendation re-ranking models according to the preference weights in a dynamic manner, avoiding the costs of re-training the models. The key of CMR is using policy hypernetworks to generate a part of the parameters in the re-ranking models. A new re-ranking model based on AE is proposed which offers an end-to-end joint optimal solution for complex business-oriented tasks. Offline and online experiments on Taobao app demonstrated the effectiveness of the proposed CMR framework and the re-ranking model. 

\begin{acks}
This work was funded by the National Key R\&D Program of China (2019YFE0198200), the National Natural Science Foundation of China (61872338, 62006234, 61832017), the Fundamental Research Funds for the Central Universities, and the Research Funds of Renmin University of China (23XNKJ13), Intelligent Social Governance Interdisciplinary Platform, Major Innovation \& Planning Interdisciplinary Platform for the ``Double-First Class'' Initiative, Renmin University of China,
and Beijing Outstanding Young Scientist
Program NO. BJJWZYJH012019100020098. 
The work was partially done at Beijing Key Laboratory of Big Data Management and Analysis Methods.
This work was supported by Alibaba Group through Alibaba Innovative Research Program.
\end{acks}

\bibliographystyle{ACM-Reference-Format}
\balance
\bibliography{ref}

\end{document}